\documentclass{article}

\usepackage{PRIMEarxiv}

\usepackage[utf8]{inputenc} 
\usepackage[T1]{fontenc}    
\usepackage{hyperref}       
\usepackage{url}            
\usepackage{booktabs}       
\usepackage{amsfonts}       
\usepackage{nicefrac}       
\usepackage{microtype}      
\usepackage{lipsum}
\usepackage{fancyhdr}       
\usepackage{graphicx}       
\graphicspath{{media/}}     
\usepackage{enumitem}
\usepackage{subcaption}

\title{Ah, that's the great puzzle: On the Quest of a Holistic Understanding of the Harms of Recommender Systems on Children
\thanks{\textit{\underline{Cite as}}: 
Robin Ungruh and Maria Soledad Pera. 2024. \textit{Ah, that’s the great puzzle: On the Quest of a Holistic Understanding of the Harms of Recommender Systems on Children.} Presented at \textit{Designing for Children’s Digital Well-
being: A Research, Policy and Practice Agenda (DCDW ’24), co-located with ACM IDC 2024}} 
}

\author{
  Robin Ungruh \\
  Delft University of Technology \\
  Delft, The Netherlands \\
  \texttt{R.Ungruh@tudelft.nl} \\
   \And
  Maria Soledad Pera \\
  Delft University of Technology \\
  Delft, The Netherlands \\
  \texttt{M.S.Pera@tudelft.nl} \\
}

\begin{document}
\maketitle

\begin{abstract}
    Children come across various media items online, many of which are selected by recommender systems (RS) primarily designed for adults. The specific nature of the content selected by RS to display on online platforms used by children---although not necessarily targeting them as a user base---remains largely unknown. This raises questions about whether such content is appropriate given children's vulnerable stages of development and the potential risks to their well-being.  
    In this position paper, we reflect on the relationship between RS and children, emphasizing the possible adverse effects of the content this user group might be exposed to online.  
    As a step towards fostering safer interactions for children in online environments, we advocate for researchers, practitioners, and policymakers to undertake a more comprehensive examination of the impact of RS on children---one focused on harms. This would result in a more holistic understanding that could inform the design and deployment of strategies that would better suit children's needs and preferences while actively mitigating the potential harm posed by RS; acknowledging that identifying and addressing these harms is complex and multifaceted.
\end{abstract}


\begin{figure}
\centering
\includegraphics[width=\linewidth]{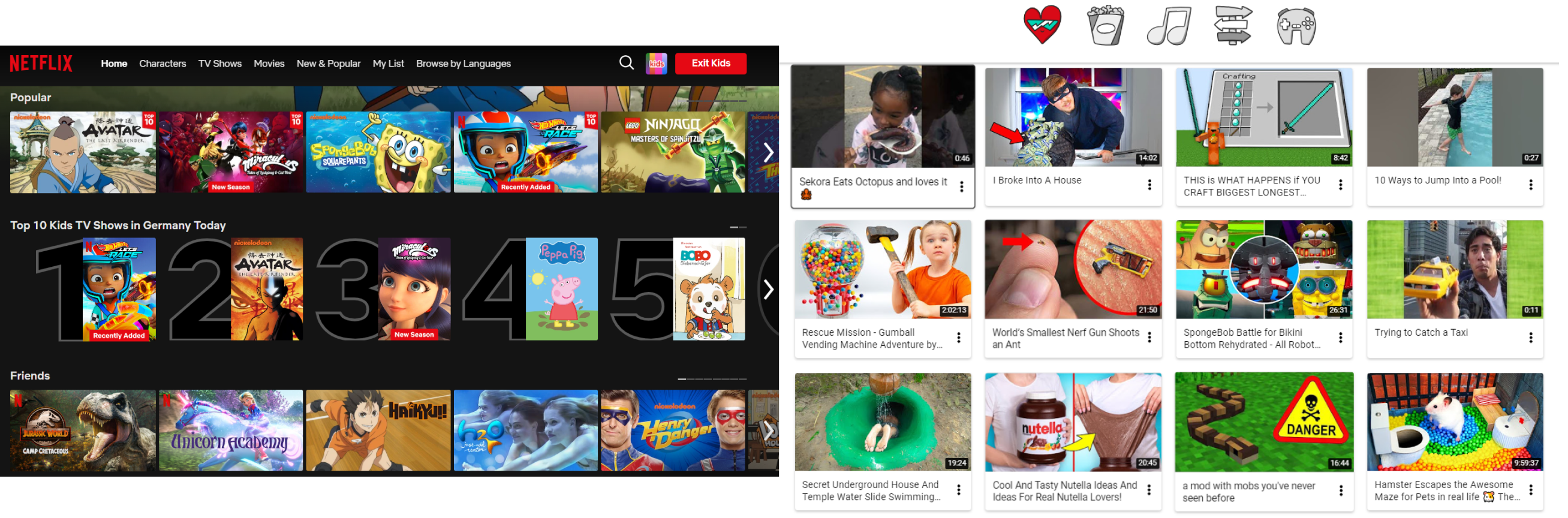}
\caption{Screenshots of \textit{Netflix}'s home page on a Kids profile (on the left) and the \textit{YouTube Kids} recommendations tab (on the right). \textit{Netflix}'s content is mainly curated and annotated with human-generated maturity ratings. \textit{YouTube} does not provide maturity ratings and consists mainly of user-generated content.}
\label{fig: teaser}
\end{figure}

\section{Down the Rabbit Hole: Children's Exposure to Online Media} 

In today's digital world, children\footnote{In this manuscript, whenever we mention children, we mean individuals aged up to 18, as per UNICEF's definition of a child \cite{UNICEF}. Given the focus of this work, we mostly refer to those who independently access online platforms.} spend a significant portion of their time online for entertainment, learning, or simply seeking information. A third of the children worldwide have internet access \cite{unicef2020many}, with significantly higher rates observed in countries across Asia, Europe, the Pacific, and the Americas \cite{unicef2020many, NCES2023Childrens, ofcom2023children}.  Reports show that children use the internet to engage with online platforms on a daily basis, consuming a wide range of content 
\cite{ofcom2023children, PRC2024teens, KMM2023Kinder}. Video-sharing platforms and social media sites are particularly popular among them, with \textit{YouTube}, \textit{YouTube Kids}, and \textit{TikTok} emerging as favorites among younger demographics along with other streaming services including \textit{Netflix} or \textit{Disney+} \cite{ofcom2023children, radesky2020young, ofcom2022children, PRC2024teens, KMM2023Kinder}. In addition to video consumption, listening to music stands out as one of children's favorite activities, with online platforms being the primary means of access. Similarly, while reading is still associated with physical books, online reading has also become a common activity for children \cite{ofcom2023children}.
 

Among commercial (streaming) platforms, there are those that primarily offer a curated catalog of items, e.g., songs, shows, and movies, either through production or licensing agreements \cite{gomez2015netflix}. This would be the case for \textit{Netflix}, \textit{Paramount+}, \textit{Disney+}, or \textit{Spotify}, to name a few. At the same time, there are platforms like \textit{YouTube} and \textit{TikTok} that enable users to upload their own content, leading to a continuously growing, uncurated catalog of items. 
Regardless of their source, these items regularly make their way to children---whether they are actively seeking them or not, see Figure \ref{fig: teaser}---which prompts the \textit{question of what kinds of content are children actually exposed to on a regular basis?} Given the potential impact of the content children consume on their development and perceptions, especially during their formative years, reflecting on these matters becomes even more urgent.

\section{The Red Queen: The Role of  Recommender Systems on Media Exposure and Potential Harms} 

As a common component of online platforms, including those frequently accessed by children, recommender systems (\textbf{RS})  play a vital role in selecting the content prominently featured in platforms' interfaces. As such, they serve as a primary means for children to discover content. 
RS examine the extensive collection of available content and strategically choose the items that best fit each user or the majority of consumers \cite{ricci2021recommender}. For \textit{Netflix}, recommendations are at the `core of [their] business'' \cite{gomez2015netflix}, relying on personalized and non-personalized strategies to produce top lists of items to showcase---and this is not the only (streaming and/or e-commerce) platform for which this statement applies \cite{jannach2019measuring, deldjoo2020recommender}. 

The prevalence of RS is also apparent among applications frequently used by these young users; for instance, \textit{YouTube Kids} \cite{ofcom2023children}  restricts the search function for children as a standard option, thus making recommendations the main way for this user group to discover videos. 
The prominence of RS in today's digital landscape highlights the fact that children will inevitably encounter RS decisions, 
i.e., items explicitly selected by RS, whether they actively seek RS assistance or not. Children become ``consumers'' of such systems, using them to access information, products, entertainment, etc. \cite{ekstrand2024not}. RS influence what children are exposed to and interact with during their online experiences. 
RS, however, are primarily designed with adult users in mind. Traditionally, RS leverage the interactions, behaviors, and preferences of adult users to promote related and often personalized content to them, aiming to enhance user experience and engagement within this demographic \cite{ricci2010mobile, ricci2021recommender, jannach2022value}. As children do not have the same requirements, needs, and preferences as the ``typical'' adult users \cite{gomez2021evaluating, ekstrand2017challenges, anuyah2019need, pera2023children, landoni2024good}, one wonders the extent to which such systems are directly applicable to children.
%

As a distinctive user group, children have unique characteristics, such as in-development cognitive abilities \cite{anuyah2019need} and comprehension skills \cite{gomez2021evaluating}, as well as preferences for different musical genres \cite{schedl2019online}. Furthermore, it is essential to recognize that children do not constitute a homogeneous user group distinct from adult users; instead, there are variations among children themselves. Comprehension and expression skills vary for different age groups \cite{feldman2004piaget}. So can interests, skills, and needs 
\cite{gomez2021evaluating, murgia2019seven, pera2023children, milton2021infinity, landoni2024good, schedl2019online}, particularly in light of traits such as the cultural background of individual users~\cite{pera2023children}.

Existing research highlights the role of RS in personalizing recommendations to support children's needs and their development \cite{gomez2021evaluating, milton2019here,charisi2022artificial}. This is evident in their ability to facilitate access to a diverse array of content and learning resources \cite{gomez2021evaluating, charisi2022artificial, pera2014automating}, while preventing information overload. By tailoring the content presented to a child, RS can also align resources with the child's skills, needs, and goals \cite{gomez2021evaluating, charisi2022artificial, ekstrand2017challenges, milton2019here}. For instance, books can be tailored to match a child's preferences \cite{milton2020don,pera2014automating} and cater to their reading capabilities and comprehension skills \cite{ng2023read, pera2013read}. 
Despite RS attempts to support children by explicitly accounting for distinctive traits, existing (academic and commercial) strategies often overlook the vulnerabilities of this population. 
During childhood development, individuals' abilities and experiences continuously evolve \cite{levin2011child}. Key characteristics, including knowledge, attitudes, and beliefs, are shaped by their observations of the world and others, especially those they admire \cite{bandura1977social}. Certain representations in media can be perceived as authentic representations of the world, with relatable characters often serving as influential role models \cite{bandura2009social}. Hence, what is presented to children, can have an immense impact on their development.

Their susceptibility to being influenced by consumed media \cite{bandura2009social} makes children vulnerable to potentially harmful content, impacting their future beliefs, attitudes, and behaviors. 
Beyond explicit and clearly identifiable harms like violence and explicit or sexual imagery \cite{albertson2018impacts, huesmann2003longitudinal}, RS might present many subtler yet also potentially detrimental types of content. Stereotypes and biased narratives \cite{raj2021pink, bigler2006developmental}, misinformation \cite{howard2021digital}, or hate-speech \cite{anuyah2020empirical}, though less obvious, can have a major impact on children's perceptions, attitudes, and behaviors during a vulnerable, developmental phase \cite{bigler2006developmental, levin2011child, bandura2009social}. 
Stereotypes are built from real-life representations in the world \cite{bigler2006developmental} and by perceiving the explicit labeling of groups and attitudes towards others \cite{bigler2007developmental}. During childhood, many characteristics of the children's personality, like values, social identity, and core beliefs, are formed \cite{bigler2006developmental}. Thus, negative stereotypes have the potential to foster prejudice, negative beliefs, and attitudes, as well as discriminatory behavior toward others later in life. RS could potentially contribute to this harm by influencing children's future beliefs, such as their perceptions of gender roles, by presenting stereotypical content \cite{seitz2020effects, kneeskern2022examining}. Research has demonstrated that biased and stereotypical narratives have the potential to influence children's beliefs and attitudes towards the representations of social groups depicted therein \cite{block2022exposure}. Moreover, misinformation is commonly recommended and even amplified by RS \cite{srba2023auditing, fernandez2021analysing}. A 2018 report \cite{NLT2018fake} shows that children struggle to judge whether a story consists of fake news and a majority of teachers believe that false information is harmful to children’s well-being. 
Thus, misinformation presented to children by RS might not be judged properly by them and could be adopted as true beliefs \cite{bandura2009social}.

Ultimately, it is not enough for RS to suggest relevant items that match children's needs and preferences. It is critical to minimize risks and prioritize children's well-being by ensuring that their interactions with RS are safe and supportive \cite{charisi2022artificial, dignum2020policy, howard2021digital}. 

\section{Through the Looking Glass: The Need for a Holistic Exploration} 
\begin{figure}
    \centering
    \includegraphics[width=.4\linewidth]{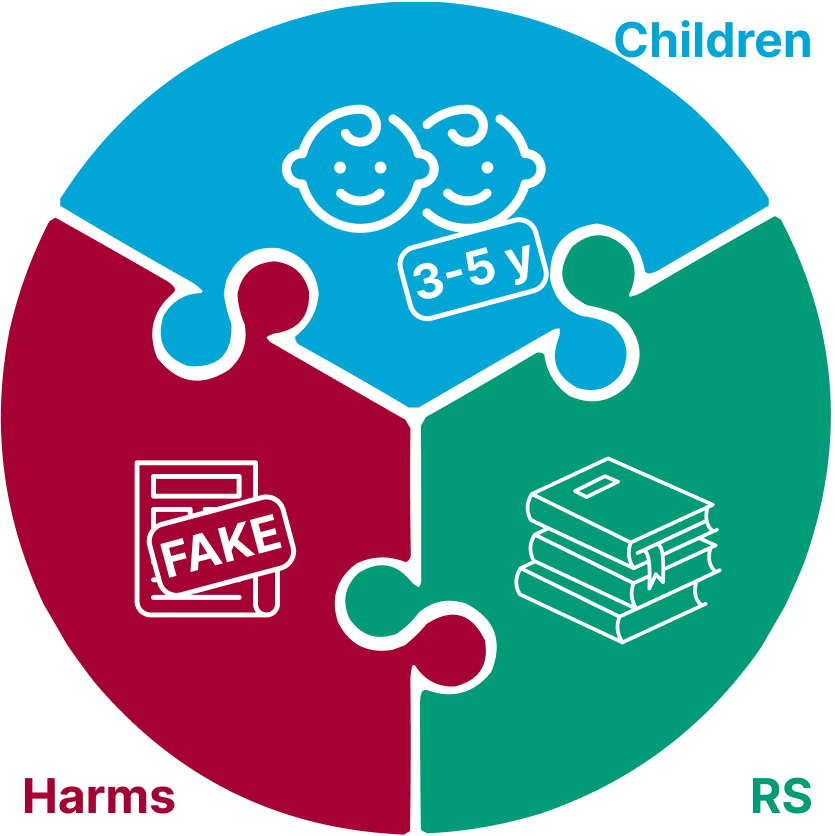}
    \caption{Visualization of the holistic view that interconnects three 3 factors for analysis: Children, RS (domain and/or strategy), and Harms. In this specific example, the holistic view focuses on scrutinizing the problem explicitly for a group of \textit{young} children, \textit{book} RS, and \textit{misinformation} as the harm.}
    \label{fig: holisticView}
\end{figure}

Due to RS' omnipresence in children's online experiences and their largely autonomous operations, concerns arise about the potential tendency of RS to present children with items that are inappropriate or harmful.
And yet, the actual effects of RS in promoting harmful content to children are largely unexplored, highlighting the need for further exploration and scrutiny. 

Expanding our understanding of the extent to which RS contribute to children's exposure to harmful content is not only necessary to inform the suggestion generation process so that it aims for safe interactions, but it also becomes a must to audit existing and new systems for harm exposure.
Researchers, designers, practitioners, and policymakers must be well-informed about the potential harms that RS could pose to children. Understanding this should prompt reflection on the impact of their decisions, thereby influencing future RS development, implementation, and deployment.


To properly gauge the risks posed by RS, we need a comprehensive understanding of their current effects on children. This requires undertaking a holistic perspective that considers and acknowledges the characteristics of $3$ factors simultaneously: \textbf{children}, \textbf{RS}, and \textbf{harms}. These factors can be viewed as puzzle pieces that, when assembled, give us a realistic picture of their interrelation in the digital landscape (see Figure \ref{fig: holisticView}). Examining these pieces in isolation will not provide a complete understanding of the complexities at play. Instead, they need to be examined in the context of one another to grasp their relationships and resulting effects. 

In practice, gaining an understanding of the factors' connection is not an easy task; it cannot be explained by a flat perspective that considers each factor in isolation. This is because each factor is comprised of various facets that influence their interplay. Children are not a homogeneous group; they have varying vulnerabilities and needs. Therefore, the same recommended item might be harmful to some children while being appropriate for others. Similarly, different harms, such as violent depictions or stereotypical representations, may have diverse impacts on children. Moreover, RS operate across different domains and make use of a wide range of algorithms and strategies, which can influence the content ultimately suggested, and indirectly, the potential harm it can propagate among children.

The multifaceted nature of each factor, as well as their complex relationships, necessitates a layered perspective for 
addressing the question of \textit{when a certain RS poses what harm to what user group in which context}. It is only when highlighting the relationship between a clearly defined subgroup of children, a specific domain or strategy of RS, and a particular harm, that we can gain insights into their interplay and the emerging negative impacts on targeted user groups. 
Consider the puzzle pieces depicted in Figure \ref{fig: holisticView}. By exploring the relationship between children aged 3 to 5, misinformation, and book recommendations, we can gain insights into how information pollution impacts young children's perceptions and beliefs \cite{landoni2023does}. However, the picture changes entirely when we alter one puzzle piece. Switching from book to movie recommendations might affect how the same misinformation is perceived. 
Since visual content is more easily comprehended \cite{crawshaw2020children}, misinformation might be grasped more readily, leading children to adopt false beliefs. Similarly, different age groups might be impacted to varying degrees by misinformation encountered in books. Older children tend to develop critical thinking skills and seek out diverse viewpoints \cite{lai2011critical}. Consequently, misinformation might not be as easily accepted, and false beliefs could be more likely to be rejected.
This scenario helps illustrate the complexity of the factors at hand, but subgroups are even more fine-grained: Children cannot be solely delineated by age groups, nor can RS be clearly segmented into different domains, or harms divided into broad effects. Each factor has more distinct characteristics that require nuanced consideration and inspection.

Let us think about the puzzle pieces associated with the children's factor. The interplay between children, RS, and harms depends, among other factors, on the unique characteristics of children. While age serves as a foundational factor in understanding what recommendations may be appropriate or harmful for them \cite{feldman2004piaget}, other individual characteristics also influence their interactions with content. Aspects such as background, past experiences, literacy skills, and cognitive abilities influence how children perceive and evaluate content. These characteristics can lead to varying perceptions of recommended content, with what is harmful to one child potentially not perceived as harmful by another. For example, while one child might easily identify misinformation or stereotypes in news articles and reflect on them, another child might struggle due to their limited experience, thus leading to the adoption of false or stereotypical beliefs \cite{bandura2009social}.
Another aspect of individual differences among children is their response to violence in media, which varies based on their previous experiences, personal characteristics (such as age, gender, and level of aggressiveness), and social environment \cite{anderson2003influence}. Furthermore, the degree of identification with characters engaging in violent behavior also plays a crucial role. According to Bandura\cite{bandura2009social}, identification with such characters can lead children to internalize their beliefs and behaviors, potentially resulting in the imitation of violent patterns \cite{anderson2003influence}.

When looking into the puzzle pieces related to the RS factor, it becomes apparent that the domain in which RS operate influences the potential harm posed by suggested content to children. Descriptions or imagery that may appear harmful in one domain can have markedly different impacts when encountered in another. 
For instance, discussions often center around the potentially greater harm of violent content in video games compared to movies, attributed to the active role players assume in games where they virtually inflict harm on others \cite{demirtacs2022violent}. However, the negative effects of violence in video games remain heavily debated \cite{ferguson2015angry}. While considering the domain of RS offers a high-level perspective, a more nuanced examination must also scrutinize the underlying algorithms and techniques employed to suggest potentially harmful content, hence probing when items are recommended that pose risks to children.

Perhaps the most complex of the puzzle pieces is the one associated with the harm factor, as there is consensus, no one-size-fits-all when it comes to defining and (automatically) detecting harms \cite{charisi2022artificial}. Consider, for instance, maturity ratings. At first glance, they seem to be suitable indicators of an item's appropriateness for specific user groups. Streaming platforms like \textit{Netflix}\footnote{\url{https://help.netflix.com/en/node/2064}} and \textit{Disney+}\footnote{\url{https://help.disneyplus.com/nl-NL/article/disneyplus-en-jp-content-ratings}} typically provide these ratings, which are generated through human assessment based on the platform's guidelines or by external organizations such as the Motion Picture Association (MPA)\footnote{\url{https://www.motionpictures.org/}} and TV parental guidelines\footnote{\url{https://movielabs.com/md/ratings/v2.3/html/US_TVPG_Ratings.html}}. The maturity rating rules of MPA include aspects like profane language, violence, nudity, sex, substance abuse, and the theme of the movie.
The validity of maturity ratings to measure the actual harm posed to children, however, is heavily debated \cite{waguespack2011ratings}. They often focus on explicit and clearly identifiable types of harm like violence, profanity, and sexual content \cite{thompson2004violence}, potentially overlooking more subtle harms like stereotypes or misinformation. In addition, they focus solely on the children's ages, although other individual differences between children might be crucial in assessing how harmful certain media is to them. Hence, while maturity ratings can provide an initial indication of the potential harm of recommended items posed to children, their severity for individuals cannot be captured by them. 
Other domains, such as books, music, or social media, lack clear guidelines for capturing the potential harm posed by their content. This is especially true for social media and video streaming platforms, which primarily consist of user-generated content. Manual appraisal of harm in these domains is often infeasible due to the sheer volume of content, demonstrating, once again, the nuanced nature of this challenge. 

Our exploration of the intricate interplay of children, RS, and harms is at a very early stage. Due to the challenging nature of avoiding harm for diverse user groups, there is much to uncover \cite{ekstrand2024not}. The challenges presented in this work do not serve as a comprehensive overview of all characteristics and the challenges involved. Instead, they serve as a starting point intended to stimulate future discussions. 

\section{Guided by the Cheshire Cat: Promoting Children's Well-Being}
Throughout this manuscript, we have reviewed the potential harms children might face in their online experiences frequently shaped by RS' suggestions. We centered on the need for a holistic perspective---one that recognizes the nuances of the interplay between RS, children's unique characteristics, and the diverse forms of harm---to help uncover a practical understanding of the risks children face as they wander around today's digital wonderland. 

We call for efforts to build a common understanding of the potential harms posed by RS to young children and adolescents in the digital landscape. 
Moving forward, it is essential to involve stakeholders: Researchers, practitioners, designers, developers, and policymakers must collaborate closely to deepen the understanding of the nuanced facets of RS interactions and their impact on children \cite{charisi2022artificial}. Through multi-domain, multidisciplinary efforts, relevant puzzle pieces can be identified and how they fit in with other pieces can be explored. The resulting pictures will facilitate the collective development of guidelines and frameworks that guide the responsible design, development, assessment, and deployment of RS for online platforms. It is in the hands of these stakeholders to ensure that RS not only enhance the user experience but also prioritize users' well-being, starting with safeguarding children from potential harm during their online experience.

\bibliographystyle{unsrt}  
\bibliography{references}

\end{document}